# The Gibbs Paradox

**Simon Saunders**

Faculty of Philosophy, University of Oxford, Oxford OX2 6GG, UK; simon.saunders@philosophy.ox.ac.uk

**Abstract:** The Gibbs Paradox is essentially a set of open questions as to how sameness of gases or fluids (or masses, more generally) are to be treated in thermodynamics and statistical mechanics. They have a variety of answers, some restricted to quantum theory (there is no classical solution), some to classical theory (the quantum case is different). The solution offered here applies to both in equal measure, and is based on the concept of particle indistinguishability (in the classical case, Gibbs' notion of 'generic phase'). Correctly understood, it is the elimination of sequence position as a labelling device, where sequences enter at the level of the tensor (or Cartesian) product of one-particle state spaces. In both cases it amounts to passing to the quotient space under permutations. 'Distinguishability', in the sense in which it is usually used in classical statistical mechanics, is a mathematically convenient, but physically muddled, fiction.

## 1. Introduction

The Gibbs paradox is usually broken down into two puzzles:

(i) Why is the entropy of the mixing of two gases independent of their degree of similarity—and only zero when the gases are the same? (the *discontinuity* puzzle).
(ii) How, in classical statistical mechanics, can an extensive entropy function be defined? (the *extensivity* puzzle).

To these we add a third, which in one form or another was highlighted in all the early discussions of the paradox:

(iii) How can there *not* be an entropy of mixing, even for samples of the same gas, in statistical mechanics, classical or quantum?—because surely the particles of the two gases undergo much the same microscopic motions on mixing, be they exactly alike or only approximately similar (the *microrealism* puzzle).

The latter is, in part, a question about the physical interpretation of the entropy quite generally.

The three puzzles are far from independent. Thus, if there is always an entropy of mixing, even for identical gases in answer to (iii), there is no discontinuity puzzle (i); and since, in that case, the entropy is not extensive, there is no extensivity puzzle (ii) either. Answers to (i) and (ii) are frequently silent on (iii). Further, each can be approached in ways that differ in classical and quantum theory. As sich, the Gibbs Paradox is complicated.

Nevertheless, we argue all three can be coherently solved in a way that takes the same form in classical and quantum theories, leading to considerable simplifications. The key concept is particle indistinguishability. Although introduced by Gibbs in a purely classical setting [1], it was subsequently annexed to quantum theory—to the point that, in classical theory, the idea has routinely been dismissed as unintelligible [2–5]. Where it has been defended, it has been interpreted in instrumentalist terms [6], or in terms of the classical limit of quantum statistical mechanics [7], or as a property of certain probability distributions [4]. These defences are perfectly adequate so far as they go, but here we take the concept further, to apply literally, at the microscopic level, to classical particle motions realistically conceived.

Section 2 is introductory, and reviews the two well-known solutions. Section 3 is on the concepts of particle identity and indistinguishability (mostly) in classical statistical mechanics; it concludes with a sketch of a solution to (i), the discontinuity puzzle. Section 4 is on (iii), the microrealism puzzle. Technical complications will be kept to a minimum. The system studied throughout is the simplest possible using the simplest tools: the ideal gas, completely degenerate with respect to energy, using the Boltzmann definition of the entropy.

## 2. Solutions

*2.1. Thermodynamics and the Discontinuity Puzzle*

Consider a volume $V_A$ of a gas composed of $N_A$ particles in region *A*, and a volume $V_B$ of a gas composed of $N_B$ particles in region *B*. Suppose they are at the same temperature and pressure and in thermal contact with a reservoir at temperature *T* (see Figure 1, the 'Gibbs set-up'), so that $N_A/V_A = N_B/V_B$.

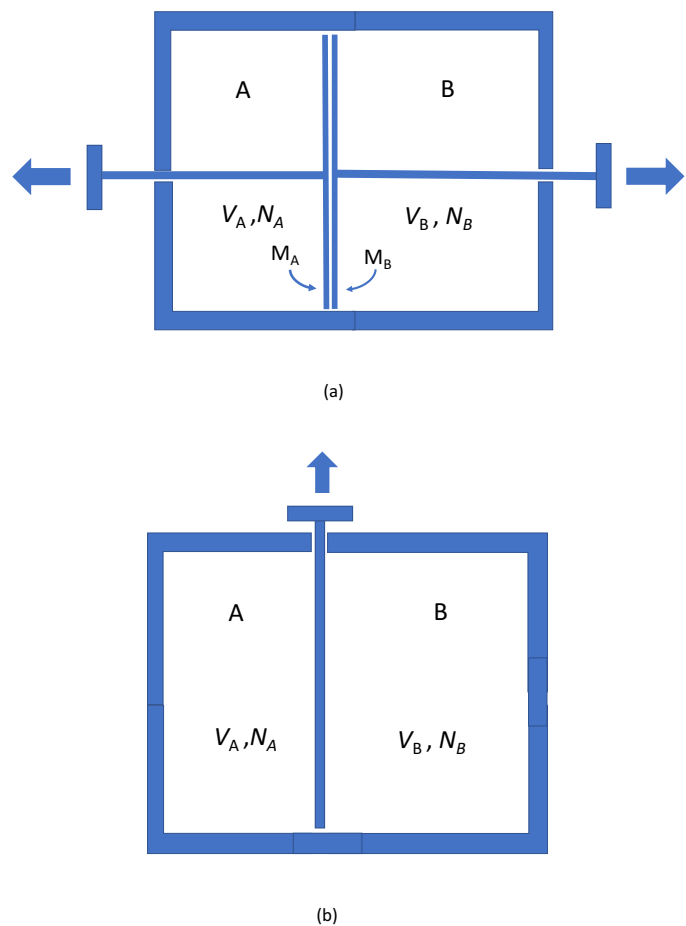

(a)

(b)

**Figure 1.** The Gibbs setup. In (**a**) the membrane $M_A$ is permeable to *A*, impermeable to *B*, whilst $M_B$ is permeable to *B*, impermeable to *A*; the pistons are allowed to expand; In (**b**) the gases are the same and a partition is removed. The pressures and temperatures in both chambers are the same.

Let the gases in *A* and *B* be distinct from each other, in the sense that they can be separated by pistons faced by membranes $M_A$ and $M_B$, where $M_A$ is permeable to gas *A* but impermeable to gas *B*, and vice versa for $M_B$. (Figure 1a). Let them slowly expand under the partial pressures of the two gases, doing work. The process is reversible, so from the work done and the equation of state the entropy increase can be calculated directly. It is (setting Boltzmann's constant equal to unity) shown by:

$$(N_A + N_B)\ln(V_A + V_B) - (N_A \ln V_A + N_B \ln V_B). \tag{1}$$

This entropy change (the entropy of mixing) is the same, however, like or unlike the two gases, so long as they are not the same.

Suppose now the gases in *A* and *B* are the same. Then no membranes of the required kind exist, and the Gibbs setup is as in Figure 1b, consisting of a single partition that is slowly removed. No work need be performed, the process is isothermal, the heat flow is zero; if the process is reversible (it seems that it is) the entropy change is zero.

The same conclusion follows from extensivity of the entropy function (given that the entropy scales with the size of homogeneous systems, along with particle number, mass, volume, and energy). The total equilibrium entropy of the two gases before the partition is removed is the sum of the two taken separately (by additivity); the total equilibrium entropy after the partition is removed is the sum of the entropy of the two sub-volumes (by extensivity); the two are the same. The change in the total entropy is zero.

How *similar* do two samples of gas have to be for this conclusion to follow? This is the discontinuity puzzle, as stated, for example, by Denbigh and Redhead [8] (p. 284):

> The entropy of mixing has the same value...however alike are the two substances, but suddenly collapses to zero when they are the same. It is the absence of any `warning' of the impending catastrophe, as the substances are made more and more similar, which is the truly paradoxical feature.

A natural response is to leave the question to the experimenter—with an attendant down-playing of an objective meaning to the entropy function. In the words of van Kampen [3] (p. 307):

> Whether such a process is reversible or not depends on how discriminating the observer is. The expression for the entropy depends on whether or not he is able and willing to distinguish between the molecules *A* and *B*. This is a paradox only for those who attach more physical reality to the entropy than is implied by its definition.

Similar remarks were made by Maxwell in his classic statement of the paradox (although he did not call it that by that name). He spoke of 'dissipated energy', defined as work that could have been gained if the gases were mixed in a reversible way (thus, he meant entropy):

> Now, when we say that two gases are the same, we mean that we cannot separate the one from the other by any known reaction. It is not probable, but it is possible, that two gases derived from different sources but hitherto regarded to be the same, may hereafter be found to be different, and that a method be discovered for separating them by a reversible process. If this should happen, the process of inter-diffusion that we had formerly supposed not to be an instance of dissipation of energy would now be recognised as such an instance. [9] (p. 646).

Gibbs, himself, when he first considered the entropy of mixing a year or two earlier, went even further:

> We might also imagine the case of two gases which should be absolutely identical in all their properties (sensible and molecular) which come into play while they exist as gases either pure or mixed with each other, but which should differ in respect to their attractions between their atoms and the atoms of some other substances, and therefore in their tendency to combine with such substances. In the mixture of such gases by diffusion an increase in entropy would take place, although the process of mixture, dynamically considered, might be absolutely identical in its minutest details (even with respect to the precise path of each atom) with processes which might take place without any increase in entropy. In such respects, entropy stands strongly contrasted with energy. [10] (p. 167)

It is undeniable that whether or not there is an entropy of mixing of two kinds of gas or fluids depends not just on the actual process, whereby the gases are mixed, but on other processes (perhaps, even, all *possible* processes). Entropy change, in the context of mixing, has a comparative dimension. However, whether that licenses the subjectivist interpretation of the entropy that Maxwell went on to draw is far from clear:

> It follows from this that the idea of dissipation of energy depends on our knowledge. Dissipated energy is energy which we cannot lay hold of and direct at pleasure, such as the energy of the confused agitation of molecules which we call heat. Now, confusion, like the correlative term order, is not a property of material things in themselves, but only in relation to the mind that perceives them. [9] (p. 646).

Gibbs [1,10] spoke rather of the entropy as defined by 'sensible qualities' (thermodynamic macrostates); Jaynes [11] of the entropy of a microstate as defined by a reference classs, a macrostate—so that one and the same microstate might have different entropies, depending on the macrostate associated with it. Van Kampen [3] best encapsulates the pragmatic tradition; it has recently been championed by Dieks and his collaborators [5,12–14].

How are experimentalists supposed to go about discriminating among molecules? Here one might think questions of sameness, or even the identity of molecules, and their dynamical interactions, might have something to do with it. However, both van Kampen and Jaynes set themselves against these kinds of ideas (calling them 'mystical' [3] (p. 309), 'irrelevancies' [11] (p. 6)) (ideas that had, however, been defended by Maxwell, for whom particles of a given chemical kind must be thought of as exactly identical and imperishable [15] (p. 254)). The microrealism puzzle was not addressed. What did remain a difficulty, in this approach, is (ii), the extensivity puzzle.

*2.2. Extensivity*

For simplicity, we use the Boltzmann definition of the equilibrium entropy. Denote the one-particle phase space by $\Gamma$, which we suppose includes the specification of the state-independent properties of the particle, like mass and charge. The phase space for $N$ identical particles is then $\Gamma^N = \Gamma \times \ldots \times \Gamma$ ($N$ factors in all). The volume of the equilibrium macrostate is of the form $(f(\alpha)V)^N$, where $\alpha$ stands for the intensive variables, yielding the equilibrium entropy:

$$S = N \ln f(\alpha) + N \ln$$

The second term on the RHS spoils extensivity, but it appears to be forced: if each of the $N$ particles can be anywhere in the spatial volume $V$, independent of the location of any other, the available phase space volume as defined by the Lebesgue measure must be proportional to $V^N$.

To give a simple combinatorial model (convenient for the later comparison with the quantum approach), suppose that all the particles have the same energy, so that the region of the one-particle phase space that we are interested in is of the form $\Gamma_V = [p, p + dp] \times V$. Let $\Gamma_V$ be fine-grained into $C$ cells each of equal volume $\tau$. Then there are $C^N$ different ways of independently distributing $N$ particles over $C$ cells, where each distribution has equal phase space volume $\tau^N$ (or equal probability – I shall use the former, but nothing hangs on the distinction). As before, the entropy cannot be extensive.

This was the form of Gibbs' paradox on its first naming, as first raised by Neumann [16] in 1891 and then by Duhem the year after [17] (see [18] for more on history):

> In a recent and very important writing, a good part of it is devoted to the definition [of a gaseous mixture according to Gibbs], Mr. Carl Neumann points to a paradoxical consequence of this definition. This paradox, which must have stricken the mind of anyone interested in these questions and which, in particular, was examined by Mr. J. W. Gibbs, is the following:

> *If we apply the formulas relative to the mixture of two gases to the case when the two gases are identical, we may be driven to absurd consequences.*

The absurdity lies in an entropy of mixing even for samples of the same gas; the unwanted conclusion that the entropy cannot be extensive. Indeed, if the volume measure is $V^N$, the entropy of mixing Equation (1) follows immediately. By additivity, the total initial entropy is:

$$S_A + S_B = \ln V_A^{N_A} + \ln V_B^{N_B}$$

whereas after the partition is removed, the entropy for the system $A \cup B$ is:

$$S_{A \cup B} = \ln(V_A + V_B)^{N_A + N_B}.$$

The difference is (1).

A simple solution is to divide the volume measure $V^N$ by $N!$ (where, in the Stirling approximation, $\ln N! \approx N \ln N - N$), a factor introduced by hand by both Gibbs and Boltzmann with no comment or justification. Following the challenge laid down by Neumann and Duhem (and in the very title of Wiedeburg's essay two years later [19], 'das Gibbs'sche paradoxen'), Gibbs was surely aware of it. He offered an answer of sort in his last major work, *Elementary Principles of Statistical Mechanics*, completed in the spring of 1901. The division by *N*! was interpreted in terms of the use of 'generic phases' (rather than 'specific phases', as hitherto) [1] (pp. 187–189)—the use of the quotient space of phase space under the permutation group. However, his only explicit justification for the move was that it gave the right answer:

> Suppose a valve is now opened, making a communication between the chambers. We do not regard this as making any change in the entropy, although the masses of gas or liquid diffuse into one another, and although the same process of diffusion would increase the entropy if the masses of fluids were different. It is evident, therefore, that it is equilibrium with respect to generic phases, and not with respect to specific, with which we have to do in the evaluation of entropy, and therefore that we must use the average [over the quotient space] and not [over phase space] as the equivalent of entropy, except in the thermodynamics of bodies in which the number of molecules of the various kinds is constant. [1] (pp. 206–207).

Few found this adequate. However, the puzzle was, anyway, soon lost in the undertow of the coming tsunami that was the discovery of the quantum. Gibbs' notion of generic phase was endorsed by Planck, but thereby associated with obscurities in Planck's own writings on entropy and the quantum (and condemned as such by Ehrenfest and Einstein [18]). It has since found few defenders. The new quantum statistics, replacing Boltzmann's, in the limit of dilute and high-temperature gases, contained the needed correction. In the decades that followed the missing N! was widely seen as evidence of the inadequacy of classical ideas, one of many shadows of the quantum.

There is, however, a notable alternative approach, which is to embrace Boltzmann's counting methods and volume measure and to follow them to their logical conclusion—under the premise that the *total* number of particles does not change (for any non-extensive function of this number will cancel on going to differences in the total entropy; and entropy differences are all that can measured). The entropy of subsystems, meanwhile, able to exchange particles with one another, may yet be extensive.

The idea was first introduced by Ehrenfest and Trkal [20] in connection with disassociated gases, which cool to give equilibrium distributions of different kinds of molecules. It has subsequently taken a number of forms [3,12,21] usually motivated by the idea that the only circumstances in which the dependence of the entropy on particle number is even *defined* are those in which particle number can actually be varied—when the system of interest can exchange particles with a particle reservoir. Computing the total phase-space volume, as before, the entropy is not extensive, but in accordance with this philosophy it does not have to be: the dependence of the entropy on the total particle number would only be defined were there an open channel enabling particle exchange with a further, and still larger, reservoir which, ex hypothesi, is not in place. However, for the open subsystem the entropy is extensive.

To see this, consider a system of N particles in volume V connected by an open channel with a particle reservoir of $N^*$ identical particles in volume $V^*$. The equilibrium macrostate (again assuming complete degeneracy in the energy) is not just the Cartesian product of the two phase space volumes $V_{V^*}^{N^*}$ and $V_V^N$, for there will be $(N^* + N)!/N^*!N!$ distinct ways of drawing N particles from the particle reservoir. The total count of available microstates should be the product of the three expressions:

$$S = \ln\left[\frac{(N^* + N)!}{N^*!N!}V^{*N^*}V^N\right] = \ln\frac{(N^* + N)!}{N^*!} + lnV^{*N^*} + \ln\frac{V^N}{N!} \quad (2)$$

In the limit of the Stirling approximation, the last term, for the entropy of the subsystem, is extensive.

As applied to the Gibbs setup, suppose there are two subsystems of interest (of volumes $V_A$, $V_B$ summing to V, and so on for the numbers of particles), initially in communication with the reservoir of $N^*$ particles. Then the initial total equilibrium entropy is:

$$S = S^* + S_A + S_B = ln\left[\frac{(N^* + N)!}{N^*!N!}V^{*N^*}\frac{N!}{N_A!N_B!}V^{N_A}V^{N_B}\right]$$
$$= \ln\frac{(N^* + N)!}{N^*!} + lnV^{*N^*} + \ln\frac{V_A^{N_A}}{N_A!} + \ln\frac{V_B^{N_B}}{N_B!} \quad (3)$$

If now the channel to the larger reservoir is closed, we suppose the equilibrium entropy is unchanged. We then have the Gibbs setup, Figure 1b. After the partition between A and B is removed, the volume measure is the same as before but for replacement of the term $\frac{N!}{N_A!N_B!}V^{N_A}V^{N_B}$ in Equation (3) by $(V_A + V_B)^{N_A+N_B}$, yielding for the equilibrium entropy:

$$S = S^* + S_{A \cup B} = \ln\frac{(N^* + N)!}{N^*!} + lnV^{*N^*} + \ln\frac{(V_A + V_B)^{N_A+N_B}}{(N_A + N_B)!} \quad (4)$$

The non-extensive factors in Equations (3) and (4) are exactly the same, so do not contribute to the entropy change. The difference between the remaining (extensive) expressions on the RHS of Equations (3) and (4) vanish in the Stirling approximation. (The argument as it stands is flawed, if only because if A and B can initially exchange particles with the larger reservoir, then they can initially exchange particles with each other. For further discussion, see [21,22,23].)

Call this the *distinguishability approach* to the extensivity puzzle. Particles are treated as distinguishable, in the sense that states that differ by the exchange of a particle in the reservoir with a particle in region V are counted as distinct, no matter that the particles are in all relevant statistical mechanical senses the same (for otherwise it would matter as to which N of the $N + N^*$ particles are in V, or which $N_A$ of the N particles are in $V_A$, etc.) (this sits uncomfortably with the supposedly pragmatic, instrumental philosophy favoured by van Kampen and Dieks).

*2.3. The Quantum Approach*

The other standard solution to Gibbs' paradox (and the one to be found in most textbooks) is to appeal to quantum mechanics. For simplicity, consider the semi-classical treatment, in which the fine-graining of the one-particle state space is determined by Planck's constant (so $\tau = h$). The number of ('elementary') cells C now has a physical meaning. When the particles are identical, the further essential assumption is that interchange of two or more particles leaves the microstate unchanged—the particles are treated, not as 'distinguishable', in the sense just defined, but in exactly the opposite way: states that differ by particle interchange are *not* distinct, hence, the rubric *indistinguishable*, now standard terminology in quantum statistical mechanics.

It follows that microstates ('Planck distributions') are fully specified by the number of particles in each elementary cell, without regard as to which particles are in which cell. Let these non-negative ('occupation') numbers be $n_1,..,n_C$, subject to the constraint:

$$\sum_{k=1}^{C} n_k = N \tag{5}$$

To determine the number of these distributions, consider sequences of $N + C - 1$ symbols, composed of $N$ symbols 'x' (one for each particle), and $C - 1$ symbols '|' (to denote the $C$ cells, counting the number of x's before the first | as $n_1$). There are $(N + C - 1)!$ permutations of the $N + C - 1$ symbols by sequence position, but not all of them yield distinct sequences: each sequence recurs $N!$ times (for permutations of the x's among themselves), and $(C - 1)!$ times (for permutations of the |'s among themselves). The number of distinct sequences is therefore [24]:

$$\frac{(C + N - 1)!}{(C - 1)! \, N!} \tag{6}$$

This expression was found by Planck, working backwards from the black-body spectral distribution law, in turn obtained by interpolating between the Rayleigh-Jeans distribution (valid at low frequencies) and the Wien distribution (valid at high frequencies). He interpreted it as the number of ways of distributing $N$ 'energy quanta' over $C$ cells ('resonators'), for radiation of frequency $\nu$, to be obtained by dividing the total energy by $h\nu$—and, thus, did Planck's constant make its first appearance. Einstein's 'light quantum' hypothesis led to the Wien distribution instead (because based on the volume measure $C^N$), as shown by Ehrenfest [25] in 1911. That same year Natanson traced the difference to the indistinguishability of particles (they were 'undistinguishably alike' [26] (p. 136), to be distributed over 'distinguishable receptacles', yielding Equation (6); but the receptacles were not clearly identified as specifying the state-dependent properties of their incumbents. He muddied the waters accordingly, adding 'were each of the [indistinguishable particles] separately sensible to us, the conditions of the case would be profoundly modified' [26] (p.136), yielding the count $C^N$ instead. ('Sensible perception', evidently, will depend on state-dependent properties, as well as state-independent ones.) Equation (6) has a significance beyond the semi-classical treatment: in terms of Hilbert space, it is the dimension of the totally symmetrised sub-space of $\mathcal{H}^N$, where $\mathcal{H}$ has dimension $C$.

To see more clearly the departure from the classical case, consider again the combinatorics argument leading to the result $C^N$. The following is an identity in number theory:

$$\sum_{\substack{\{n_k\} \text{ s.t.} \\ \Sigma_{k=1}^C n_k = N}} \frac{N!}{n_1! \dots n_C!} = C^N.$$

The sum is over all sets of occupation numbers satisfying Equation (5), as before (so the number of terms in the summand is given by Equation (6)), but each term (for each distinct Planck distribution) is weighted by the factor $N!/n_1! \dots n_C!$, corres/Nponding to the number of distinct ways ('Boltzmann distributions') of dividing $N$ particles so that $n_1$ are in the first cell, ..., and $n_C$ in the last. The particles are, thus, being treated as distinguishable, in our technical sense. If Boltzmann distributions have equal phase space volume, or probability, then Planck distributions do not, and vice versa, save in the limit in which all the occupation numbers are 0s and 1s. In the latter limit $C \gg N$, and Equation (6) goes over to the corrected volume measure $C^N/N!$.

This much speaks in favour of the equiprobability of Boltzmann distributions (and, hence, distinguishability): only then is the assignment of each of the $N$ particles, made sequentially, one after the other, among the $C$ cells, statistically independent of each other. If Planck distributions are equiprobable instead, still taking the assignment sequentially, the probability that the *k*th particle is assigned to a given cell increases with the number already assigned to it. Indistinguishable quantum particles are not statistically independent in this sense. (On the other hand, the whole idea of building up to a microstate sequentially, assigning the particles one by one, may be mistaken. The $N$ particles may be better thought of as assigned all together, with the microstate supervening globally.)

If statistical independence in this sense is a mark of the classical, so too is the equiprobabilty of Boltzmann distributions, hence, distinguishability: that lent support to the view that division by $N!$

can only be explained by the quantum. The idea (we take it) is that there *are* no classical gases or substances, but that since real gases are quantum mechanical systems, treatable, to a greater or lesser accuracy, by semi-classical methods, that go over, in the dilute limit, to the classical expression for the entropy, differing only by the needed correction, the division by *N*! is explained.

The extensivity puzzle (ii) is thereby solved. It may well be true that the dependence of the entropy on particle number can only actually be *measured* if the particle number is allowed to change, but opening a channel to a particle reservoir is not what introduces the needed N! factor; that factor is already there (the puzzle, as such, does not arise). Additionally, unlike in the distinguishability approach, there is no constraint on total particle number.

The discontinuity puzzle (i) also appears solved (but here appearances are deceptive): quantum theory not only implies a discretisation of the energies of bound states, at the level of atomic and molecular structure, it also explains how there can be an *exact* identity of particles at all (with respect to their state-independent properties)—because they are excitations of a single quantum field. The (anti-)symmetrization is, moreover, truly built in; the only way of arriving at a particle representation of a quantum field at all, is in terms of a state space (Fock space), built up from totally (anti-)symmetrised states. Wiedeburg's conclusion in 1894 in light of the Gibbs paradox was prophetic:

> The paradoxical consequences [of the mixing-entropy formula] start to occur only when we follow Gibbs in imagining gases that are infinitely little different from each other in every respect and thus conceive the case of identical gases as the continuous limit of the general case of different gases. On the contrary, we may well conclude that finite differences of the properties belong to the essence of what we call matter. [19] (p. 697).

Realism at this level, however—essentially concerning the spectrum of the energy operator for bound states and the simple harmonic oscillator—does not extend straightforwardly to a solution to (iii), the microrealism puzzle. That puzzle, recall, is that at least in the classical case, on any realistic perspective, there patently *should* be diffusion from the gas in *A* into *B*, and vice versa. Why not an entropy of mixing, even when the gases are the same? For the quantum approach to repudiate the entire question of microrealism turns it into another doctrine altogether (instrumentalism, say); but then, from a microrealist perspective, how *does* a quantum gas diffuse? Schrödinger, famously, suggested that for a quantum gas there *is* no real diffusion, but he only hinted at an argument as to why [27] (p. 61):

> It was a famous paradox pointed out for the first time by W. Gibbs, that the same increase of entropy must not be taken into account, when the two molecules are of the same gas, although (according to naive gas-theoretical views) diffusion takes place then too, but unnoticeably to us, because all the particles are alike. The modern view [of quantum mechanics] solves this paradox by declaring that in the second case there is no real diffusion, because exchange between like particles is not a real event—if it were, we should have to take account of it statistically. It has always been believed that Gibbs' paradox embodied profound thought. That it was intimately linked up with something so important and entirely new [as quantum mechanics] could hardly be foreseen.

Is it true that in quantum mechanics 'exchange between like particles is not a real event'?—and what does this have to do with diffusion? However, on questions like these, and on micro-realism more generally, there is no consensus in quantum theory, and is reason, if possible, to pursue the puzzle in classical terms.

## 3. Reconsidering Indistinguishability

The approaches just sketched do not have to stand opposed. The distinguishability approach may plausibly apply to the statistical mechanics of macroscopic objects, like stars in stellar nebula, and sufficiently complex microscopic systems, like colloid particles in suspensions, where the interchange of particles surely does make for a physical difference; and there the pragmatic stance of van Kampen and Dieks seems unproblematic. As a matter of course, on the quantum approach, whenever the dependence of the equilibrium entropy function of a system on particle number is actually to be measured, there had better be an open channel allowing a change in particle number, or equivalent.

However, agree on this much and you confront the obvious question: how does a difference in *scale* (stars), or in *complexity* (stars, colloid particles) break permutation symmetry, exactly? If the distinguishability approach fails in the case of ordinary gases of simple molecules, what replaces it?

The quantum approach is more right than the distinguishability approach, but it needs to handle these exceptions, and explain how intrinsic distinctions can arise at all—and embrace parity of treatment of identical particles in classical, as in quantum, statistical mechanics. To that end, we need a better understanding of what indistinguishability really means, quantum and classical.

*3.1. Indistinguishabilty and Sequence-Position*

The standard objection to particle indistinguishability in classical statistical mechanics is that 'classical particles can always be distinguished by their trajectories' (e.g., [5] (p. 373)), and even 'classical indistinguishable particles have no trajectories' (they can only have probability distributions) [4] (p. 7). Evidently this means distinguishability with respect to their *state*-dependent properties, whereas indistinguishability as we are using it is about state-*independent* properties.

Trajectories per se are irrelevant to indistinguishability. The point is most simply made in quantum mechanics: *quantum* particles can, sometimes, be distinguished by their trajectories, without ceasing to be indistinguishable (in the state-independent sense) [28] (pp. 199–200), [29] (pp. 358–359). If identical, the one-particle Hilbert space $\mathcal{H}$ for each particle is identical. Consider any set of $N$ pairwise-orthogonal one-particle states in $\mathcal{H}$ $\{|\varphi_a\rangle, |\varphi_b\rangle, \ldots, |\varphi_c\rangle\}$. Let $\Pi^N$ be the permutation group acting on the $N$ symbols $a, b, \ldots c$, so that for $\pi \in \Pi^N$, $\pi(a), \pi(b), \ldots, \pi(c)$ is a sequence of the same symbols, but in a diferernt order. Define the state $|\Psi\rangle \in \mathcal{H}_S^N$, where $\mathcal{H}_S^N$ is the symmetrised sub-space of $\mathcal{H}^N = \mathcal{H} \otimes \ldots \otimes \mathcal{H}$ ($N$ factors in all), as:

$$|\Psi\rangle = \frac{1}{\sqrt{N}} \sum_{\pi \in \Pi^N} |\varphi_{\pi(a)}\rangle \otimes |\varphi_{\pi(b)}\rangle \otimes \ldots \otimes |\varphi_{\pi(c)}\rangle. \tag{7}$$

States of the form Equation (7) describe bosons. They span $\mathcal{H}_S^N$, the quantum state-space for $N$ bosons. So long as there are no repetitions, they are in 1:1 correspondence with the unordered sets $\{|\varphi_a\rangle, |\varphi_b\rangle, \ldots, |\varphi_c\rangle\}$.

Suppose now that the particles are non-interacting and prepared in a state of the form Equation (7); then they remain in a state of this form. If non-interacting, the Hamiltonian $\widehat{H}$ is a sum of one-particle Hamiltonians $\hat{h}$, all identical (since $\widehat{H}$ is permutation invariant), generating the unitary evolution:

$$|\Psi\rangle \to e^{i\widehat{H}t}|\Psi\rangle = \frac{1}{\sqrt{N}} \sum_{\pi \in \Pi^N} \widehat{U}_t|\varphi_a\rangle \otimes \widehat{U}_t|\varphi_{\pi(b)}\rangle \otimes \ldots \otimes \widehat{U}_t|\varphi_c\rangle$$

where $\widehat{U}_t = e^{i\hat{h}t}$. Each of the $N$ one-particle states $|\varphi_a\rangle \in \mathcal{H}$ etc., initially orthogonal to all the rest, remains orthogonal at each time, and traces out a definite orbit $U_t|\varphi_a\rangle$ in $\mathcal{H}$.

The argument can easily be elaborated. Let each of $|\varphi_a\rangle$, $|\varphi_b\rangle$, ..., $|\varphi_c\rangle$ (call them $a$, $b$, .., $c$ for short) be well-localised in phase space, well-separated from each other, and sufficiently massive to ensure that they remain well-localised and well-separated over the timescale of interest. Given an (external) time-dependent potential function, the trajectories thus defined can be as varied as is desired, each distinguished from all of the others. In short, we obtain a good approximation to $N$ non-intersecting trajectories in one-particle phase space. Yet they are described by a totally symmetrised state—and, therefore, as indistinguishable particles.

If the having or not-having of trajectories is irrelevant to indistinguishability (in the state-independent sense), might it have something to do with entanglement, necessarily introduced by symmetrisation? However, insofar as the state is entangled *only* for this reason that seems unlikely. States of the form Equation (7) fail to satisfy any of the important desiderata for entanglement [30–32]. Genuine (or so-called 'GMW') entanglement involves the superposition of states of this form.

Another response (in light of the selfsame example of quantum trajectories) is to conclude that (anti-)symmetrisation of the state *has nothing to do* with the notion of distinguishability—or not as relevant to the Gibbs paradox [13]. The latter, in this view, concerns state-dependent properties. Yet (anti-)symmetrisation of the state is all-important to quantum departures from classical statistics (to obtain Bose-Einstein or Fermi-Dirac statistics). Where particles cannot be distinguished by their state-dependent properties, they yet obey Maxwell-Boltzmann statistics, unless they are (anti-)symmetrised. Lack of statistical dependence clearly hinges on (anti-)symmetrisation, and just as clearly bears on the Gibbs paradox (a point we shall come back to in Section 4.3)

We conclude rather that *a, b, …,* and *c* are distinguished as one-particle *states*, but what they are states *of*, as specified by their state-independent properties, are exactly alike—given that the particles are identical. Permutations of particles with respect to *a, b, …* (as to which particle is *a*, which particle is *b*….) do not yield distinct states of affairs; likewise for permutations with respect to the trajectories.

Why then introduce names for particles in the first place? This is because names come with sequence position: the order in the N-particle Hilbert space $\mathcal{H}^N = \mathcal{H} \otimes … \otimes \mathcal{H}$. Names cannot help but have mathematical significance, so long as sequences are used, and outside of statistical mechanics, sequence position usually has a clear physical significance, with each system, entering into the tensor product, having its own distinctive degrees of freedom and coupling constants. The Hamiltonian, correspondingly, is by no means permutation invariant in its action on such a product space. However, in *statistical mechanics*, in dealing with $10^{20}$ + particles, they had better all have at least approximately the same mass and coupling constants, if the equations are to be defined at all. The one-particle state spaces are then the same, and the Hamiltonian will be blind to sequence position, and so be permutation invariant. (Anti-)symmetrisation of the state and symmetrisation of dynamical variables ensures that no use can be made of sequence position to label particles.

Use of sequence-positions as names for identical particles has been called 'factorism' (my thanks to Jeremy Butterfield and Adam Caulton for this way of putting it); particle indistinguishability, then, is anti-factorism. It can be applied equally to classical mechanics, where the counterpart is sequence position in the Cartesian product state-space $\Gamma^N = \Gamma \times \Gamma \times … \times \Gamma$ (N-factors in all), and sequence-position in ordered N-tuples $q = \langle q_a, q_b, …, q_c \rangle \in \Gamma^N$. The ordering is eliminated, as in the quantum case, but now for *N* pairwise distinct points (this the analogue of orthogonality, where we suppose particles are impenetrable, so that the condition is preserved in time), by passing to unordered sets of one-particle states [33] (pp. 176–177). As a differentiable manifold, the state space is the quotient of $\Gamma^N$ under the permutation group $\Pi^N$ with the action:

$$\pi: \langle q_a, q_b, …, q_c \rangle \to \langle q_{\pi(a)}, q_{\pi(b)}, …, q_{\pi(c)} \rangle \in \Gamma^N. \qquad (8)$$

The topology is the image of open sets in $\mathbb{R}^N$ under the quotient. The resulting 'reduced' state space is $\gamma^N = \Gamma^N/\Pi$. It is the space of generic phases, in Gibbs' terminology (whereas $\Gamma^N$ is the space of specific phases [1] (pp. 187–188)). The quotient map defines a surjection from the smooth kinematically possible motions $\sigma \subset \Gamma^N$ to smooth motions in reduced state space, denote $\sigma^g \subset \gamma^N$. (Since $\gamma^1$ and $\Gamma = \Gamma^1$ are isomorphic, we denote the one-particle phase-space $\gamma$ when speaking of motions of indistinguishable particles.)

Evidently for any permutation $\pi \in \Pi^N$ and any $q \in \Gamma^N$, the reduced point $q^g = \pi(q)^g \in \gamma^N$ represents an unordered set of points in $\gamma$. For the action of permutations on particle trajectories, consider first the unreduced case. A smooth curve $\sigma \subset \Gamma^N$ parametrised by $\lambda \in \mathbb{R}$ is a continuous map $: \lambda \to \sigma(\lambda) \in \Gamma^N$, representing an ordered *N*-tuple of one-particle trajectories in $\Gamma$, as defined by the smoothly varying (and never intersecting) *N*-tuple of points at each instant of time. However, then, for any permutation $\pi$, the curve $\pi(\sigma): \lambda \to \pi(\sigma(\lambda)) \in \Gamma^N$ *also* represents an *N*-tuple of one-particle trajectories in $\Gamma$, indeed, the very same trajectories, differing only in their factor-position (or in their names --names as factor-positions). In the reduced state-space this is eliminated: there is just the one curve $\sigma^g = \pi(\sigma)^g \subset \gamma^N$, representing *N* trajectories in $\gamma$, with no factor-positions.

There is this key difference between classical and quantum. In the quantum mechanics of *N* particles, in the general case, a state of N particles does not define any *one* set of *N* one-particle states. In the face of genuine (GMW-) entanglement (and real physical particles interact, and interactions lead to GMW-entanglement), this resource is not available. However, classically, there is always this resource; it is alway*s* possible to speak of one-particle states (points) $q_a, q_b, \ldots, q_c$ in $\gamma$, and indeed to identify particles by their position and momentum at a time (so speak of *a*, *b*, … and *c* at a given time). Doing so is already to pass to the quotient space, under permutations. Were it not for GMW-entanglement, it would be the same in quantum mechanics, and we could talk of one-particle states directly, with no need to talk of particles in (anti-)symmetrised states

There is another important difference between the classical and quantum statistical mechanics of identical particles that makes their similarity much harder to see. Use of the reduced state space is all but compulsory in the quantum case, but it is hardly ever used classically. Why? The answer is that, classically, there is an easy correspondence between integrals of permutation-invariant functions on the reduced phase space and on the unreduced space, as pointed out by Gibbs [1] (p. 188).

To illustrate the correspondence in the simplest case, consider a multiple integral over the domain $a \leq x_1 \leq \cdots \leq x_N \leq b \subseteq \mathbb{R}^N$, so that there can be no repetition of arguments assigned these variables. Let $f: \mathbb{R}^N \to \mathbb{R}$ be permutation invariant. Then:

$$\int_a^b \int_a^{x_{N-1}} \ldots \int_a^{x_1} f(x, x_1 \ldots, x_{N-1}) dx dx_1 \ldots dx_{N-1}$$
$$= \frac{1}{N!} \int_a^b \int_a^b \ldots \int_a^b f(x_1, \ldots, x_N) dx_1 dx_2 \ldots dx_N \qquad (9)$$

The left-hand side, extended now to 6*N*–dimensions, is the integral of *f* over $\gamma_{[a,b]}^N$, the right-hand side the integral of *f* over $\Gamma_{[a,b]}^N$, divided by *N*!. The only difference is the needed correction—and that can be traced to quantum mechanics instead, as we have seen. Integrals as on the right of Equation (9) are, needless to say, much easier to perform than those on the left. There is no gain, and considerable pain, in doing analysis on the reduced state space; better do it on $\Gamma^N$ instead, with judicious insertions of factors in *N*! as needed. Particle indistinguishability, in classical statistical mechanics, becomes all but invisible. The reason it becomes visible in equilibrium *quantum* theory is because of the lower bound to the size of cells (by Planck's constant) and the concomitant replacement of the continuous measure on $\gamma^N$ by the count of Planck distributions for elementary cells (or, in terms of measures on Hilbert-space, the dimensionality of the symmetrised Hilbert space). That makes for a straightforwardly measurable difference in the statistics away from the dilute limit. (Indistinguishability is needed: the quantum statistical mechanics of *distinguishable* particles obeys Maxwell-Boltzmann statistics. For further discussion, see [28,29].)

Outside of the classical limit, there is no *general* correspondence of the form Equation (9) for expectation values calculated with respect to the unreduced and reduced Hilbert space (although there is for certain equilibrium states). Thus, any gain is limited. However, there is no pain in working with the reduced space: (functional) analysis on the reduced space is just as easy as on the unreduced space. The root of the asymmetry lies in the topology (of the spaces as topological spaces). $\Gamma^N$ is homeomorphic to $\mathbb{R}^N$, but not so $\gamma^N$ (roughly speaking, sets bounded by lines and planes $x_i = x_j, x_i = x_j = x_k$, etc., invariant under permutations are open sets in $\gamma^N$, but not in $\mathbb{R}^N$). In contrast, the reduced Hilbert space $\mathcal{H}_S^N$ has exactly the same topology as the unreduced space: it is a closed subspace of $\mathcal{H}^N$, in the norm topology, a complex Hilbert space isomorphic to any other of the same dimension.

*3.2. Permutations as Active Transformations*

Particle permutations on $\Gamma^N$ as defined by Equation (8) act as identities in the reduced phase space $\gamma^N$. If the latter is to be taken seriously as the space of microstates, in a fully realist way, permutations in this sense cannot represent real physical changes. However, permutations surely *can* be real physical processes. As Pais has graphically put it [34] (p. 63):

> Suppose I show someone two identical balls lying on a table and then ask this person to close his eyes and a few moments later to open them again. I then ask whether or not I have meanwhile switched the two balls around. He cannot tell, since the balls are identical. Yet I know the answer. If I have switched the balls, then I have been able to follow the continuous motion which brought the balls from the initial to the final configuration. This simple example illustrates Boltzmann's first axiom of classical mechanics, which says, in essence, that identical particles which cannot come infinitely close to each other can be distinguished by their initial conditions and by the continuity of their motion.

'Switching the balls' means a physical change—a continuous curve in the state space parameterised by the time. Consider, to begin with, the unreduced phase space $\Gamma^2$ of the two balls, representing two trajectories in the one-particle phase space $\Gamma$—from a state at one time $t_1$ to a state at $t_2$ arrived at by particle exchange, with space-time diagrams as shown in Figure 2. In (a) the same state is returned, whereas in (b) it is physically switched. A physical switch is a real physical process, not a mere relabelling of points or trajectories.

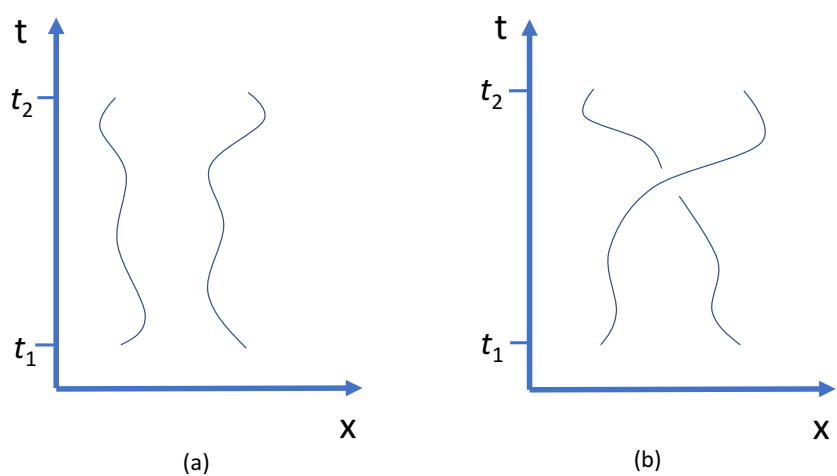

**Figure 2.** Space-time diagrams of two identical particles with the same initial and final positions and momenta. In (**a**) each particle has the same position and momentum at $t_2$ as at $t_1$; In (**b**) they are physically switched: each particle at $t_2$ has the position and momentum of the other at $t_1$.

For the generalisation to $N$ particles, for any $q \in \Gamma^N$ let $\sigma_\pi: t \to \Gamma^N$ be a smooth curve connecting $q_1$ to $q_2 = \pi(q_1)$:

$$\sigma_\pi(t_1) = q_1$$

$$\sigma_\pi(t_2) = \pi(q_1).$$

It represents $N$ trajectories in $\Gamma$, beginning in one microstate, and ending in a microstate exactly the same save that it is arrived at by the interchange of initial and final positions and momenta of two or more particles. The $N$ trajectories represent a physical switch (of course, physical switches, in practise, could never be made exact, for motions on smooth manifolds, but they are *kinematically possible*.)

Now to the point: physical switches in the reduced state space return the same state. The *open* curve $\sigma_\pi$ in $\Gamma^N$ reduces to a *closed* curve $\sigma_\pi^g$ in $\gamma^N$, for we have the sequence of identities:

$$\sigma_\pi(t_2)^g = \pi(q_1)^g = q_1^g = \sigma_\pi(t_1)^g.$$

Notwithstanding the fact that the permuted particles are individually changed, the initial and final states are the same. (The suggestion, once again, is that the microstate is defined by the state of all N particles taken collectively, rather than built up from each considered in isolation, as noted earlier.)

There is no difficulty in embracing this conclusion so long as it is hedged: if the states in question are coarse-grained, if the initial and final states are not really the same (but that for all practical purposes they can be treated as the same)—in which case, there is no particular reason why the particles should really be identical, either (a point we shall come back to). However, suppose the particles are identical and there is no more exact level of description and we are taking the theory literally: can microstates, the points in state-space at which a physical switch begins and ends, be exactly the same?

Here is the sort of principle that would rule against it: 'if the state of each of two things is changed, the state of both things together is changed' (I am grateful to a conversation with Thomas Davidson for making a case of this kind). However, that principle is not a logical truth, and again, in quantum mechanics, it is easy to construct a counter-example. Thus, consider, as before, an initial ('trivially entangled') symmetrised state of the form Equation (7) for $N = 2$, the case of two bosons:

$$|\Psi\rangle = \frac{1}{\sqrt{2}}(|\varphi_a\rangle \otimes |\varphi_b\rangle + |\varphi_b\rangle \otimes |\varphi_a\rangle). \tag{10}$$

Let the particles be non-interacting, as before, but now let the Hamiltonian generate the continuous unitary evolution from $t_1$ to $t_2$, satisfying:

$$\widehat{U}|\varphi_a\rangle = |\varphi_b\rangle; \widehat{U}|\varphi_b\rangle = |\varphi_a\rangle.$$

Then:

$$|\Psi(t_2)\rangle = \widehat{U} \otimes \widehat{U}|\Psi(t_1)\rangle = |\Psi(t_1)\rangle = |\Psi\rangle$$

and the orbit of the state is a closed curve in $\mathcal{H}_s^2$: particle $a$ turns into particle $b$, and $b$ into $a$, in the same microstate Equation (10) in which they collectively began. Let $|\varphi_a\rangle$ and $|\varphi_b\rangle$ have definite *shapes*, rather than positions and momenta, say one is square and one is round at $t_1$: then something square changes into something round, whilst something round turns into something square, ending at $t_2$ in exactly the same state in which they began. Each has changed so that nothing has changed—it is not difficult to find this paradoxical. No wonder Gibbs' idea of generic phase, taken realistically, and not just reflecting our epistemic limitations, has been found puzzling. It is perfectly consistent all the same.

Returning to Figure 2, are not the *histories* of the two states different at $t_2$?—of course; but that is a function of assuming *possible* trajectories, or *possible* dynamics. Fixing on (b), then, and the dynamics as shown in the figure: is not the history of the state at $t_2$ different from that of the state at $t_1$? That depends. Suppose the Hamiltonian is time-dependent, so the curve does not repeat. Then prior to $t_1$, to be sure, the history can be anything you please; as after $t_2$ as well. However, these are clearly not dictated by a difference in the *states* at the two times, but by differences in the dynamics at earlier and later times. For a time-independent Hamiltonian the pattern repeats; in which case not only is the history of the two-particle state the same at $t_2$ and $t_1$, so is the history of each one-particle state. There is no physical evidence, no logical inconsistency, no a priori argument, to speak against it. Indeed, on reflection, there had better *not* be such an argument, or it would tell against the standard treatment of indistinguishability in terms of Feynman path integrals (that identifies states, but now in configuration space, that differ by particle interchange, so the sum over paths includes both kinds of trajectories, with and without particle interchange). Similarly, for quantization on reduced configuration space [35].

At this point the connection with the Gibbs paradox, and specifically (iii), the micro-realism puzzle, is fairly direct. Roughly speaking, after the partition is removed (in Figure 1b), it seems that particles from $A$ can be found in $B$, and vice versa, and these are possibilities that were not present before. That is to say: additional states, over and above those available before the partition is removed, appear to be available at later times. However, those additional states differ from the ones accessible before the partition was removed only by a physical switch, a closed loop in $\gamma^N$. There are new *trajectories*, among them new closed loops, that become kinematically possible, but no new *points*, differing only by a physical switch. Figure 2 is the Gibbs paradox for two particles. If it is a bullet, bite.

We shall return to this argument in Section 4. Before that, a final piece of stage-setting is needed.

*3.3. Demarcating Properties*

There is nothing in principle to prevent us treating every physical property as a state-dependent property, so that all particles whatsoever have the same state-independent properties (namely none at all) (a speculation in physics, whether trivial [36], or by grand-unification). The same could be said of properties of macroscopic bodies, indeed, of ordinary bodies (arriving, at the end, at 'bare particulars', a speculation in philosophy). At the other extreme is the idea that all bodies, including microscopic particle, have uniquely distinct state-independent properties ('tropes', perhaps, or 'haecceities', further speculations in philosophy), by virtue of which they are distinguishable.

The truth, we suppose, lies somewhere in between. However, where is there any middle ground—how do any bodies, or particles, *start* to become distinguishable?—and we are back to (i), the discontinuity puzzle. However, it is now more clearly posed as a puzzle about how state-independent properties arise, and are used in a state-space description. The answer, from a dynamical point of view, is that they arise in a given regime, stable in time, yet salient to the dynamics. Certain degrees of freedom are effectively frozen, but variation in others remain, defining an effective state-space.

Call such properties *demarcating properties* [33]. The paradigm case is a disassociated gas, particles of which combine to form stable molecules in equilibrium at definite concentrations (the model studied by Ehrenfest and Trkal [20]). For an example with the changing particle number (that cannot be handled by the distinguishability approach), consider a plasma of neutrons at high temperatures and pressures, cooling to plasmas of protons, helium, lithium, and beryllium nuclei, and their isotopes, electrons, and antineutrinos, and then to gases and metallic vapours. At each stage the dynamics simplifies, as first stable particles and nucleons are formed, and then neutral atoms and molecules. This process of differentiation into kinds occurs when particles are confined to certain regions of state space, governed by an effective Hamiltonian, where particles (or bound states of particles) of a given kind remain indistinguishable. The idea of a *distinguishable* particle, as uniquely specified in this way, is not impossible—individual atoms can be manipulated in the laboratory—but from the point of view of *statistical* mechanics it is a limiting case.

For a toy model consider $N$ indistinguishable classical coins in a box, and suppose the coins interact elastically and are subject to gravity. At sufficiently low energies their motions are confined to horizontal and vibratory motions—the coins are never or rarely flipped, but freely move from left to right sides of the box. Each is confined to one of two regions of the single-coin phase space, $\gamma_H$ and $\gamma_T$ (where $H$ corresponds to coins landed heads-up, and $T$ for tails). Let there be $N_H$ heads-up coins and $N_T$ tails-up; then, ex hypothesis, as long as the kinetic energies remain small (the box is not violently shaken), the motions will be confined to the region $\gamma_H^{N_H} \times \gamma_T^{N_T} \subset \gamma_{H \cup T}^{N}$. The coins behave as distinct collectives, one (the heads-up) differentiated from the other (the tails-up), in a dynamically salient way (the side face-down makes a difference to the friction, say); but each a collection of indistinguishable coins. The embedding of $\gamma_H^{N_H} \times \gamma_T^{N_T}$ in $\gamma_{H \cup T}^{N}$ for $N_H = N_T = 1$ is shown in Figure 3.

Thus, as long as demarcating properties matter to the dynamics, the Hamiltonian, as a function on the product space, will not be permutation invariant. Additionally, insofar as these properties are dynamical in origin, there is every reason to think the dynamics simplifies (certain degrees of freedom are frozen out). Demarcating properties, where they exist, are part and parcel of an effective dynamics and an effective structure to state-space. Evidently similar remarks apply to quantum mechanics: thus neutrons evolve in time to states that explore a subspace of the total Hilbert space isomorphic to the (unsymmetrised) tensor product of the space of (symmetrized) states of hydrogen atoms $\mathcal{H}_S^{N_A}$ with the space of (symmetrized) states of helium atoms $\mathcal{H}_S^{N_B}$, of the form $\mathcal{H}_S^{N_A} \otimes \mathcal{H}_S^{N_B}$, or broken down further into isotopes (and likewise for lithium and beryllium). The story for the rest of the chemical elements is more complicated, involving the life-cycles of stars, but is similarly embedded in nuclear physics, whilst the fixing of chemical properties generally is in the completely different regime of atomic and molecular physics. All of this is entirely familiar, unproblematic, and deep. It is the remarkable story of the dynamical emergence of complex stable molecules not as distinguishable particles, but as natural kinds.

Returning to the Gibbs paradox, it is clearer that (i), the discontinuity puzzle, never was a puzzle about how to pass from identical particles to distinguishable particles. It was a puzzle about the differentiation of gases (each a gas of indistinguishable particles), and how such a differentiation can arise in a continuous way. The answer is that differentiation is *emergent*, arising with demarcating properties, better or worse defined, more or less robust under perturbations, changing in these respects in a continuous way. Thus, in the case of the coins initially elastically scattering at high enough energies, *H* and *T* are not demarcating properties, and there is only one kind of particle; but slowly reduce the total energy, and H and T *become* demarcating properties, and there are two kinds of coins. The transition is by degrees. If work is to be extracted on mixing, it will be more or less efficient in consequence, the emergent structure more or less robust and well-defined.

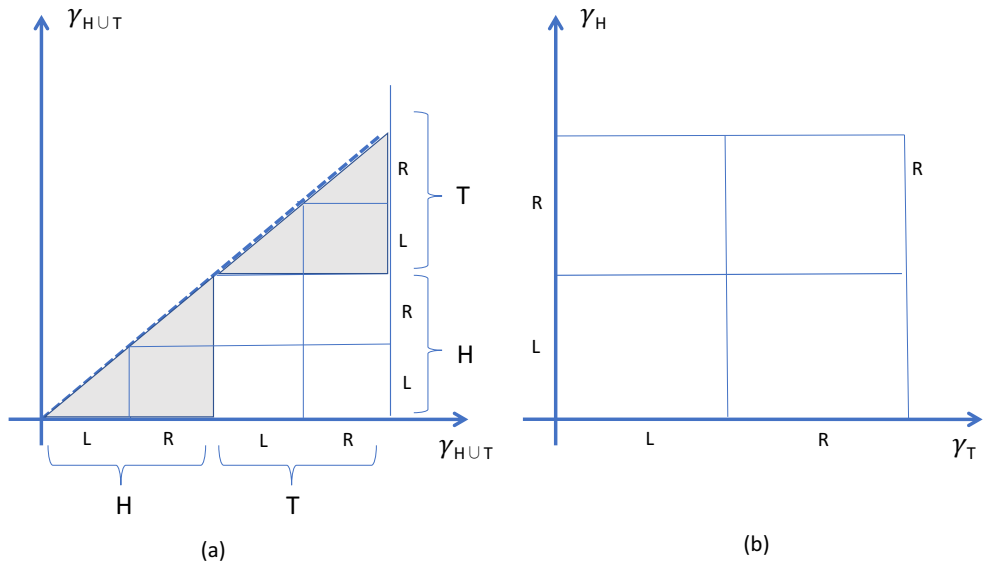

**Figure 3.** Effective state-space for identical coins coarse-grained with respect to coins that are heads-up (region *H*), tails-up (in *T*), and in the left and right sides of the container (*L* and *R*), for $N_H = N_T = 1$. The region shaded in (**a**) is inaccessible at the relevant energies. The effective available state space is isomorphic to (**b**), treating *H*-coins and *T*-coins as distinct.

Additionally, clearer is that the indistinguishability approach can, perfectly well, be applied to particles that are not really identical, with respect to their state-independent properties, at all—as witness the coins! As macroscopic bodies they, the coins in your pocket, differ in countless ways, stable in time, surely even in their state-independent properties (supposing the state-dependent properties concern only their bulk degrees of freedom). The question is only whether those differences matter to the effective dynamics. If not, they might as well be treated as indistinguishable. It is the same for the mixing of suspensions of colloid particles, suitably grouped by mass and moments of inertia. (This case deserves special consideration, and may well be a test-case for the approach favoured here. For relevant background, see [37]. Another test case is the kind of continuously-variable demarcating properties of the sort considered by von Neumann [38] and Landé [39]).) It is likewise for stars in the collision of galaxies, grouped by mass and angular momentum—or people in statistical economics models, grouped by incomes. These were the cases that were supposed to favour the distinguishability approach: they do nothing of the kind.

However, for two collections of such 'particles', all within the same group, on mixing, will there not *really* be an increase in entropy? No doubt; but as van Kampen remarked, an experimentalist who ignores it 'will not be led to any wrong results' [3] (p. 306) (by treating them as indisdstinguishable)—unless, of course, she ups her game, and finds a more accurate model and a more subtle method of mixing the two collections sensitive to a finer-grained set of state-independent properties (and indistinguishables now defined by the latter). In relation to this, in the original mixing process there is indeed an entropy of mixing. Continuing all the way so that only a single colloid particle or star remains for each set of state-independent properties, will yield a model of distinguishable particles, and a Hamiltonian with no permutation symmetries: but it will hardly be a model in *statistical* mechanics at all. It is rather the full *N*-body problem, for *N* different masses and coupling constants, in all its intractable complexity. Whether there is entropy change on mixing involves a comparison with possible physical processes whereby the gases are reversibly separated: but they had better be *thermodynamic* processes (see also the 'demon' argument of Section 4.1).

The example of the coins illustrates the Gibbs paradox in another way. For suppose their markings fade with wear, and eventually disappear altogether. Will it not remain true that the effective state space is the unshaded region of Figure 3a?—there will always have been just one coin, heads up, and just one coin, heads down, no matter that the markings have faded away entirely. The analogue, in the Gibbs paradox, is the place of origin.

## 4. The Micro-Realism Puzzle

The discontinuity puzzle (i) does not arise; (ii), the extensivity puzzle, is arguably solved. We are engaged with (iii), the micro-realism puzzle. For an early statement by Wiedeburg [19] (p. 693):

> However, if we admit the mental or even practical possibility to reversibly mix or unmix similar [gas] masses in such a way that every individually determined smallest particle is found in the same 'state,' in particular in the same position, after a complete cycle, it cannot be denied that in such a mixing process work can be won even though it does not involve any outward change.

For one much more recent: [12] (pp. 1304–1305):

> If the two gases are chemically speaking the same, the mixing will not be detectable by looking at the usual thermodynamical quantities. This is so because in thermodynamics we restrict ourselves to the consideration of coarse-grained macroscopic quantities, and this entitles us to describe the mixing of two volumes of gases of the same kind, with equal $P$ and $T$, as *reversible* with no increase in entropy. However, if we think of what happens in terms of the motions of individual atoms or molecules, the two processes (irreversible and reversible mixing) are completely similar. In other words, the qualification of the mixing process as irreversible or reversible, and the verdict that the entropy does or does not change, possesses a pragmatic dimension. It depends on what we accept as legitimate methods of discrimination; chemical differences lead to acknowledged thermodynamical entropy differences in a process of mixing, whereas mere differences in where particles come from do not.

They identify the main question: what is a 'legitimate' method of discrimination, in terms of the individual particle motions? Is a 'mental possibility' sufficient?

*4.1. Place of Origin as a Demarcating Property*

Can atoms and molecules be sorted as to their place of origin (from *A* or from *B*)? Equivalently: do the atoms from region *A* and region *B* differ in some dynamically salient property, which can be manipulated by the experimenter? In terms of Section 3.3: can place of origin function as a demarcating property?

*In general* the answer is negative. After equilibriation, the place of origin in *A* or *B* is not, in general, a useful way to arrange a coupling in the Hamiltonian, or to locate a degree of freedom obeying any simple equation, or to identify any emergent phase-space structure. In special cases—given sufficiently small numbers of particles, with the right kind of initial state and dynamics—it surely may; but if we are speaking of ordinary gases at ordinary temperatures and pressures, on the basis of everything we currently know, there is no effective dynamics for particles that came from *A*, different from that for particles that came from B, if the particles have the same state-independent properties, by means of which they could be resorted. Particles that really do have the same state-independent properties, no matter how precise the dynamics (in the regimes where they exist at all), produce no entropy in mixing. A pragmatic approach to the definition of the entropy makes sense at the level of complex systems (but favours rather the indistinguishability approach, rather than that of van Kampen et al., as illustrated by the coins); realism takes over when it comes to simple microscopic particles, whose state-independent properties really are identically the same, and halts the regress in comparison with ever more refined methods of mixing.

Against this is Wiedeburg's casual eliding of 'mental' and 'practical' possibilities. It has recently been revived in the form of a 'demon' argument [5] (p. 372):

> Figuratively speaking, think of submicroscopic computers built into the membrane that perform an ultra-rapid calculation each time a particle hits them, to see where it came from; or the proverbial demon with super-human calculational powers who stops or lets pass particles depending on their origin. In general, of course, allowing expedients of this kind may upset thermodynamical principles, in particular the second law of thermodynamics. However, in the thought experiment we propose here we make a restricted use of these unusual membranes. The idea is merely to employ them for the purpose of demonstrating that if gases are mixed and unmixed by selection on the basis of past particle trajectories and origins, as should be possible according to classical mechanics, this leads to the emergence of an entropy of mixing.

Whether or not there is an entropy of mixing of gases depends on possible dynamical processes whereby they can be separated—granted. Is there not the mere *possibility* of a membrane thus imagined sufficient to conclude there is always an entropy of mixing, even for identical gases?

However, it is noteworthy that neither Maxwell, nor Thomson, appealed to demons with these capabilities. Indeed they would not be *Maxwell* demons at all—creatures that Maxwell and Thomson took pains to insist were just like simple mechanisms (just very tiny) [40] (pp. 3–6). They are rather demons in the sense of Laplace, possessed of computational powers sufficient to trace the real-time motions of $10^{20}$-plus interacting particles through the entire process of equilibration and beyond. These are powers not super-human, but supernatural. (Notice that there is no difficulty in accommodating sorting actions of Maxwell demons which *can* be modelled as simple mechanical systems, in terms of our framework of demarcating properties. Values of certain degrees of freedom of the demon may well be used as demarcating properties, simplifying the statistical mechanical description of the process of sorting.)

Since, in reality, so far as we know, there are only quantum microscopic particles, and because, in general states they will be GMW-entangled, so that there will be no particle trajectories (putting to one side hidden-variable theories), it could be argued that, in reality, not even an array of Laplace demons could restore the original situation. That is surely true, if confined to local operations, but if the demons act in concert why not suppose, since we are granting them unlimited computational powers, that they can reverse the global quantum state, and restore the particles to each side of the partition in that way? Or, better still, give up on this line of argument altogether.

The better conclusion to be drawn is that if two samples of a gas are to be reliably separated from one another, as a matter of the local physics, then they had better *in fact* differ in some occurent demarcating property, or in some occurent stable dynamical property that can *become* salient and, hence, that can function as a demarcating property, one that an effective, local Hamiltonian can actually see. Chemical properties, shapes, and composition of molecules are prime examples of properties that are of this kind; place of origin, once the partition is removed and equilibration has occurred, is a prime example of a property *not* of this kind.

*4.2. Equilibration*

Why is this, exactly?—why do details of the past not matter to the 'effective' dynamics? The question can be posed, even, considering the gas contained in *A* in isolation (when there is no gas in region *B*).

The answer is specific to equilibrium statistical mechanics. Place of origin does not matter because we suppose that the entire region of state space consistent with the equilibrium macrostate is available, no matter that the *actual* history of the gas implies that only a tiny fraction of the macrostate will be explored. At the fine-grained level, when equilibrating with increase in entropy, a gas evolves, under the Hamiltonian flow, into an enormously fibrillated structure spreading throughout the newly-available phase-space volume. But, on pain of violating Liouville's theorem, it occupies exactly the same volume as before. The 'newly-available volume', in contrast, is much larger, corresponding to the new equilibrium macrostate with a correspondingly greater entropy.

Crucially, *so long as we move forward in time*, we do not go wrong in choosing a random microstate in this new equilibrium region for future predictions (and use a uniform probability density or volume measure over the macrostate accordingly). Any such choice is as likely to produce entropy-increasing behaviour into the future as is the *actual* microstate of the gas. The reason is that the dynamics is *forward-compatible*, to use Wallace's terminology [41]: forward evolve from $t_0$, coarse-grain -- take the average over coarse-grained regions of phase space -- forward evolve, coarse-grain, repeat, ending at some final time $t_n$: the result is the same as evolving from $t_0$ to $t_n$ and coarse-graining only once at the end. (This is a claim that needs to be—and has—been proved case by case.) The fact that any such choice is just as likely to produce entropy-increasing behaviour into the *past*, unlike the actual microstate, need not disturb us at all.

The arguments for this approach to reconciling thermodynamic irreversibility with an underlying classical reversible dynamics (essentially solving Loschmidt's Paradox), including the need for a low-entropy initial state to the universe (the 'past hypothesis'), have been widely debated, and have recently reached some consensus, at least in the philosophy of physics literature [42]). They go through more or less unchanged in quantum mechanics [43],[44] (pp. 324–360) (although this point is more contentious). We take all this as given.

We use it to conclude, in the specific case of diffusion of $N_A$ particles initially confined to $A$ into the volume $V$, that the effective state space (whether on reversible or irreversible expansion into $V$) may be taken as the full phase space volume corresponding to the spatial volume $V$, yielding an increase in the equilibrium entropy, and that this is the same whether we use the reduced or unreduced phase space. The fact that, in actuality, it is impossible that every phase space point in the equilibrium macrostate $\Gamma_V^N$ (or $\gamma_V^N$) could be explored by the gas under the actual dynamics, given that all the particles were originally confined to region $A$, will not matter in the slightest. However, if now we suppose that there are initially $N_B$ particles in $B$ as well, identical to those in $A$, to take the product of the volumes of the equilibrium states (with the partition removed) of each considered separately will lead to overcounting: of points not contained in the Hamiltonian flow from $A$ (but contained in the flow from $B$), and vice versa. In this sense the equilibration of the gas from $A$, and the gas from $B$, are not independent processes. This point repays further attention.

*4.3. Independence*

As we have seen, when the count of cells in phase space has a physical meaning, and away from the dilute, high-temperature limit, identical particles are not independently distributed. However, in the classical case, where the dilute limit $C \gg N$ can always be taken, the occupation numbers are 0s and 1s, and statistical independence is restored. In what sense, then, is the diffusion of like gases in classical theory into a common volume not independent?

Consider first the case of non-interacting but distinct gases. The initial state space for the Gibbs set-up is then $\gamma_{V_A}^{N_A} \times \gamma_{V_B}^{N_B}$, with the Hamiltonian sensitive to factor-position in this pair of factors. When the partition is removed, this factor structure remains, and the effective state space is enlarged to $\gamma_V^{N_A} \times \gamma_V^{N_B}$. The available volume has increased from:

$$\frac{C_A^{N_A}}{N_A!} \frac{C_B^{N_B}}{N_B!} \tag{11}$$

to:

$$\frac{(C_A + C_B)^{N_A}}{N_A!} \frac{(C_A + C_B)^{N_B}}{N_B!} . \tag{12}$$

The logarithm of the ratio of Equations (11) and (12) is Equation (1), the entropy of mixing (recall that $C_A/C_B = V_A/V_B$). The total entropy is the sum of the entropies of the two gases, $A$ and $B$. Each has diffused throughout $V$ as if the other were not there (because non-interacting), and each increases its entropy.

Let the gases now be identical and, as before, let them be non-interacting (so the gas is ideal). As long as the partition is in place, 'region $A$' and 'region $B$' continue to function as demarcating properties; the physical exchange of particles from $A$ with particles from $B$ is kinematically impossible, and the Hamiltonian is different for the A particles than for the B particles (the forces directed by the partition on the A particles are different from those directed on the B particles). The available phase space structure is, therefore, initially the same as for unlike gases, $\gamma_{V_A}^{N_A} \times \gamma_{V_B}^{N_B}$, and the volume of the initial equilibrium macrostate is as before, Equation (11). When the partition is removed, however, granted that 'origin in $A$' and 'origin in $B$' define no effective phase space structure, no effective Hamiltonian, and do not function as demarcating properties, the available state space is $\gamma_V^N$, and not $\gamma_V^{N_A} \times \gamma_V^{N_B}$. In place of (12) the new phase-space volume is:

$$\frac{(C_A + C_B)^{N_A+N_B}}{(N_A + N_B)!} \ . \tag{13}$$

Evaluating the logarithms of Equations (11) and (13) in the Stirling approximation, the answers are the same, and there is no entropy of mixing.

Observe that Equation (13), unlike Equation (12), cannot be written as the product of the volumes of the two equilibrium macrostates for $A$ and $B$ considered separately; the entropy of the total system is not the sum of the entropies of its two parts once the partition is removed, even when the parts—the gases from $A$ and $B$—are non-interacting. Each taken on its own, with the removal of the partition, would increase in entropy; if each diffused independently of the other, the entropy of each would increase and, therefore, the sum of the two entropies would increase as well

To understand why independence fails, suppose $C \gg N$, so the occupation numbers consist only of 0s and 1s. Microstates defined with respect to this fine-graining merely record which $N$ of the $C$ cells are occupied by single particles. Consider any one of these microstates: there are $N!/N_A!N_B!$ ways of writing it as an ordered pair of microstates for $N_A$ and $N_B$ particles, corresponding to all the ways of partitioning the $N$ occupied cells into $N_A$ cells and $N_B$ cells—equivalently, corresponding to all particle exchanges between $A$ and $B$. However, these are not distinct *microstates* (not even when reduced to points, as argued in Section 3.2; we shall come back to this in the next section).

Considering the $N_A$ particles from $A$ and the $N_B$ particles from $B$ as independently distributed after the partition is removed, is to suppose the state-space has the structure $\gamma_V^{N_A} \times \gamma_V^{N_B}$,—but that structure (or rather the structure $\gamma_{V_A}^{N_A} \times \gamma_{V_B}^{N_B}$) disappeared with the removal of the partition. Appealing to it all the same, the factor position is used to draw a distinction—of each occupied cell, whether it is of the A-type or the B-type (whether it is associated with the first factor or the second factor). However, when the particles are identical, the cells only come in one type, for that kind of particle; and, otherwise, are just fine-grained values of position and momentum.

Microstates are not built up by first distributing $N_A$ particles among the C cells, and then $N_B$, independent of the first $N_A$ (counting all the ways this can be done, in this order), no more than they are built up by assigning particles singly, each independent of the others (and counting all the ways *this* can be done, in order). In the first case the overcount is $N!/N_A!N_B!$; in the second case it is $N!$.

### 4.4. Particle Trajectories

We concluded there was no pairing of each occupied cell, after the partition is removed, with A-type or B-type; no specification of cell, by reference to place of origin of the particle that occupies it. That may be true of cells in some fine-graining, but surely if we get down to actual *points*, there is always a pairing—provided by the actual dynamics.

Thus, much must be granted (and marks a difference from quantum mechanics, in the absence of particle trajectories). However, as we saw in Section 4.2, the equilibrium entropy is a measure of all phase space points kinematically accessible, not just those in the Hamiltonian flow given the actual dynamics and initial state. The crucial question is whether there are points kinematically accessible after the shutter is released, that were not accessible before.

Consider Figure 4, depicting space-time diagrams for 'typical' kinematically possible particle motions in the Gibbs set-up from $t_1$, before the partition is removed, to $t_2$, after, in the case $N_A = 3$, $N_B = 4$. In (a), the partition remains in place throughout; in (b) it is removed at time $t'$. At $t_2$ the microstate (now a point in reduced phase space) of (a) and (b) is the same, considered as an unordered set of positions and momenta, but it has been reached in (b) in a way impossible in (a). Indeed, there are $N!/N_A!N_B!$ distinct pairings possible (35 in the case of Figure 4). Does this mean that there are 35 distinct *microstates* at $t_2$ all of them identical as unordered sets of positions and momenta? Not by virtue of any state-independent property, if the particles are identical; nor, by supposition, with respect to any state-dependent property.

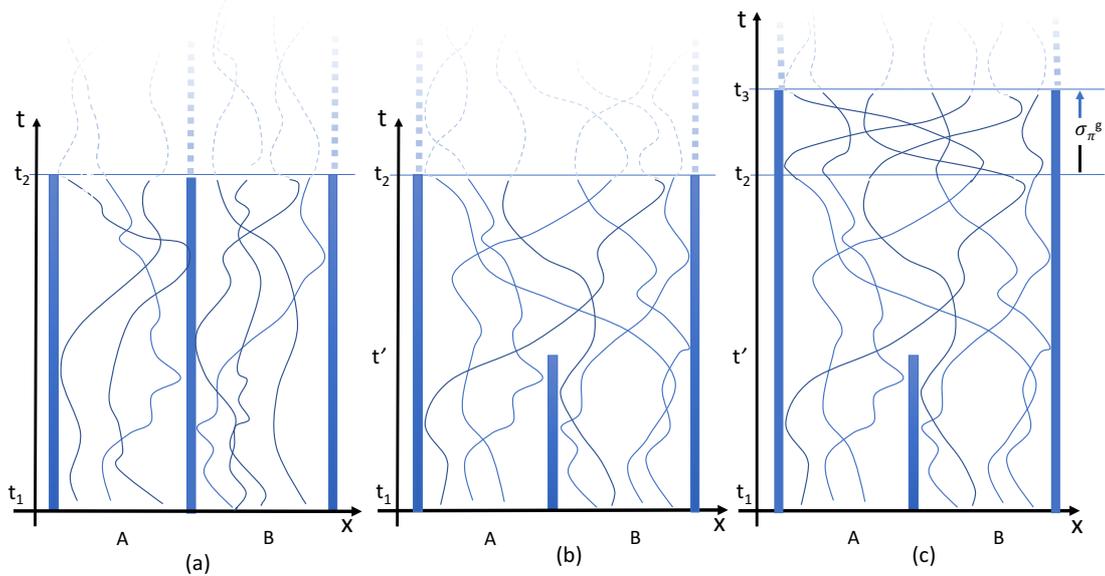

**Figure 4.** Space-time diagram of typical particle trajectories in the Gibbs set-up. In (**a**) the partition remains in place; In (**b**) the partition is removed at $t'$. The particle positions and momenta, without regard to pairings in *A* and *B*, are the same in (**a**,**b**); In (**c**) particles are physically switched between $t_2$ and $t_3$ by the curve $\sigma_\pi^g$, restoring the state at $t_3$ to the pairing in (**a**) at $t_2$.

Observe now that the two microstates at $t_2$ in Figure 4(a) and 4(b) are the beginning and end points of physical switches. In Figure 4c, the motions of (b) are extended by the curve $\sigma_\pi^g$ (or $\sigma_\pi$ in the unreduced space), where $\pi$ is the permutation restoring the pairing between initial and final states as in (a). However, as a curve in the reduced state space, physical switches, we know from Section 3.2, begin and end at the same point. The microstate at $t_3$ in Figure 4(c) is uncontroversially the same as at $t_2$ in (a) (because the pairings are the same), but are also the same as at $t_2$ in (b) (because they are connected by a physical switch); therefore, the microstates at $t_2$ in (a) and (b) are the same.

If physical switchings of identical particles produce distinct microstates there is an entropy of mixing for identical gases: if not there is none. The microrealism puzzle comes down to this.

For confirmation that this is indeed what drives the microphysical puzzle [14] (p. 742) (for $N_A = N_B = N$, $V_A = V_B = V$):

> Consider again two gas-filled chambers, each containing *N* identical particles. Before the partition is removed the number of available states per particle is *V*. After the partition has been removed, the number of available states is 2*V*. The reason is that after the partition's removal it has become possible for the particles to move to the other chamber. The doubling of the number of available microstates thus expresses a physical freedom that was not present before the partition was taken away. Trajectories in space-time have become possible from the particles' initial states to states in the other chamber. Particles from the left and right sides can physically exchange their states.

It is true that new *trajectories* become possible, but the question is whether new *microstates* become accessible. The same ambiguity is evident in the quotation from Pais. Our answer is that there are none—or not because of exchanges of particles between $A$ and $B$.

There is, however, *another* reason why new points in phase space become available, when the barrier is removed—and why, as is evident from Figure 3a, there *is* an entropy of mixing even of identical gases. There are points reachable by motions in Figure 4b, not reachable by any motion of the form Figure 4a—namely, those that differ in how many particles at $t_2$ are in region $A$, and how many in $B$. At the level of the fine-graining into $C$ cells, new occupation numbers are permitted, summing to new numbers $N_A' \neq N_A, N_B' \neq N_B$ (but still, of course, summing to $N$). Accounting for all these, the total number of possible distributions after the partition is removed is (summing over all possible values of $N_A$ and $N_B$ consistent with the constraint):

$$\sum_{N_A+N_B=N} \frac{C_A^{N_A}}{N_A!} \frac{C_B^{N_B}}{N_B!} \tag{14}$$

which is (exactly) equal to the volume given by Equation (13). The reason that Equation (14) *also* agrees with Equation (11), in the Stirling approximation (when (11) is only *one* term in the summand of Equation (14)), is because the overwhelming majority of states are those for which the number-densities are the same ($N_A/C_A = N_B/C_B$). The situation in this regard is no different from the pressure, where fluctuations in the equilibrium state are possible, but make a vanishingly small contribution to the entropy. (Thus, in Figure 4b, the trajectories are 'typical' in that the same numbers $N_A$ and $N_B$ at $t_2$ are the same as at $t_1$).

## 5. Conclusions

We may probe the quantum mechanical state space using the same technique of kinematically possible trajectories, using special states and unitary equations of the form considered in Section 3, in which case the entire argument of Section 4 goes unchanged. (However, it is hardly an adequate microrealist account of diffusion in more general states, an account in which decoherence will figure prominently.) To return to Schrödinger's remark, the 'exchange between like particles' is a perfectly possible real process (as is diffusion itself), but it is consistent with no entropy change, for it produces no new kind of event.

We are surely pushing at an open door in the quantum case. What of the classical? If the initial and final states of Figure 2b can be identified when the trajectories are those of quantum particles, there can be no *general* prohibition against the identification of states related by switches when the trajectories are those of classical particles. There is no difficulty with their mathematical identification, as shown in Section 3.1; there is no logical impediment; a metaphysical argument can surely be constructed to defeat the identification, but it is as likely to defeat itself (it is not a virtue of a speculative metaphysical argument that it contradicts our best physics)—unless, much more excitingly, it is shown how to bring it to experimental test (as Bell showed, in the case of local determinism).

Failing disconfirmation of this kind, we conclude that physical switches of identical particles return the same state, as a consequence of passing to the quotient space of N-particle state spaces under permutations, the same in classical mechanics as in the quantum case. With that there is no longer a rationale for an entropy of mixing of identical gases on the basis of microrealism, even in classical theory. No new points in phase space are explored, after the barrier is removed, other than those involved in statistical fluctuations of number densities.

The microrealism puzzle is comprehensively solved. The extensivity puzzle does not arise. The discontinuity puzzle is solved by the recognition that it is a puzzle about the differentiation of substances in terms of demarcating properties. Gases of distinguishable particles, in the traditional sense, are convenient fictions, useful by virtue of Equation (9). Where there are large numbers of complex bodies (like colloid particles or stars), insofar as they can be treated by statistical mechanical methods at all, it had better be possible to group them into kinds, each with similar state-independent properties (as goes the effective dynamics), hence, as indistinguishable within each kind. The real application of the concept of distinguishable particles is not in statistical mechanics at all: it is in the *N*-body problem, classical or quantum.

**Funding:** The support of the Leverhulme Foundation is gratefully acknowledged.

**Conflicts of Interest:** The author declares no conflict of interest.